\documentclass[conference]{IEEEtran}
\IEEEoverridecommandlockouts
\usepackage[utf8x]{inputenc}
\usepackage{cite}
\usepackage{amsmath,amssymb,amsfonts}
\usepackage{algorithmic}
\usepackage{graphicx}
\usepackage{textcomp}
\usepackage{xcolor}
\usepackage{booktabs}
\usepackage{url}

\usepackage{breakurl}
\usepackage[breaklinks]{hyperref}

\def\BibTeX{{\rm B\kern-.05em{\sc i\kern-.025em b}\kern-.08em
    T\kern-.1667em\lower.7ex\hbox{E}\kern-.125emX}}
\begin{document}

%-------------------- Otsikko ja nimet -------------------

\title{Improving Energy Expenditure Estimation in Wearables using a Heat Flux Sensor: First Observations
\thanks{© 2020 IEEE.  Personal use of this material is permitted.  Permission from IEEE must be obtained for all other uses, in any current or future media, including reprinting/republishing this material for advertising or promotional purposes, creating new collective works, for resale or redistribution to servers or lists, or reuse of any copyrighted component of this work in other works.}}

\makeatletter
\newcommand{\linebreakand}{%
  \end{@IEEEauthorhalign}
  \hfill\mbox{}\par
  \mbox{}\hfill\begin{@IEEEauthorhalign}
}
\makeatother

\author{\IEEEauthorblockN{Saku Levikari}
\IEEEauthorblockA{
\textit{LUT University}\\
Lappeenranta, Finland \\
saku.levikari@lut.fi}
\and
\IEEEauthorblockN{Antti Immonen}
\IEEEauthorblockA{
\textit{LUT University}\\
Lappeenranta, Finland \\
antti.immonen@lut.fi}
\and
\IEEEauthorblockN{Mikko Kuisma}
\IEEEauthorblockA{
\textit{LUT University}\\
Lappeenranta, Finland \\
mikko.kuisma@lut.fi}
\and

\IEEEauthorblockN{Heikki Peltonen}
\IEEEauthorblockA{
\textit{University of Jyväskylä}\\
Jyväskylä, Finland \\
heikki.peltonen@jyu.fi}
\linebreakand
\IEEEauthorblockN{Mika Silvennoinen}
\IEEEauthorblockA{
\textit{University of Jyväskylä}\\
Jyväskylä, Finland \\
mmjsilvennoinen@gmail.com}
\and
\IEEEauthorblockN{Heikki Kyröläinen}
\IEEEauthorblockA{
\textit{University of Jyväskylä}\\
Jyväskylä, Finland \\
heikki.kyrolainen@jyu.fi}
\and
\IEEEauthorblockN{Pertti Silventoinen}
\IEEEauthorblockA{
\textit{LUT University}\\
Lappeenranta, Finland \\
pertti.silventoinen@lut.fi}
\and}

\maketitle

% --------------- Dokumentti alkaa tästä ---------------------

\begin{abstract}
Wearable electronics are often used for estimating the energy expenditure of the user based on heart rate measurement. While heart rate is a good predictor of calorie consumption at high intensities, it is less precise at low intensity levels, which translates into inaccurate results when estimating daily net energy expenditure. In this study, heart rate measurement was augmented with heat flux measurement, a form of direct calorimetry. A physical exercise test on a group of 15 people showed that heat flux measurement can significantly improve the accuracy of calorie consumption estimates when used in conjunction with heart rate information and vital background parameters of the user.
\end{abstract}

\begin{IEEEkeywords}
Heat flux, wearable sensors, energy expenditure
\end{IEEEkeywords}

\section{Introduction}

The advent of wearable electronics has enabled consumers to measure their vital signs during their everyday life. Applications such as smart watches are gaining ground in measuring biometric signals from the user. Optical heart rate (HR) tracking, or photoplethysmography (PPG), is one of the most common biometric measurements in smart watches, with a market penetration of 98\% \cite{Chang}. One of the most common applications for wearable devices is to measure the user's energy expenditure (EE). This gives the wearer the ability to track their physical activity habits, which may be beneficial for weight loss and  sports performance monitoring. However, several studies have reported that the accuracy of the EE estimate in many wearables is relatively low. Recent studies have shown that the mean EE error of commercially available smart watches and activity trackers typically ranges between 25 and 50\% at rest or when performing physical activities such as walking and cycling \cite{Shcherbina_2017, Acc_of_commercial_smartwatches}.

%Sherbina et al. measured the EE estimation performance of several wearables during low-intensity physical activities such as walking and cycling, and showed that the error in the EE estimate typically ranges between 25 and 50\% \cite{Shcherbina_2017}.

One key reason for the inaccurate EE estimates is the limited number of measurements available in wearables. While heart rate has been observed to be a good indicator of EE at moderate-to-high intensity activities, the error increases during low intensity activity \cite{Jkyla_HR_EE}. This is especially problematic for smart watches and activity trackers intended for everyday use, because the resting metabolic rate typically accounts for 60-80\% of the total daily energy expenditure \cite{EE_and_aging}. Therefore, in order to achieve higher EE estimation accuracy in everyday activity monitoring, heart rate tracking should be complemented with other measurements.

During physical activity, the waste heat emitted by the human body increases. Thus, an accurate way of determining a person's EE is to use room calorimetry, which measures the entire heat output of the human body. While this technique is very impractical outside the laboratory environment, there is an alternative method to measure the waste heat output locally using a heat flux (HF) sensor. However, the heat flux measurement is strongly affected by the measurement location and ambient conditions, and thus far has remained untapped in the field of wearable technology, besides a few exceptions \cite{EE_single_site_HF}.

The hypothesis of this study was that local heat flux measurement in conjunction with heart rate tracking can improve the accuracy of energy expenditure estimate in wearable applications, namely smart watches and bracelets. To this end, physiological trials were conducted on 15 persons, and the subjects' EE was estimated using heart rate, heat flux and the combination of the two. The results show that the best overall accuracy is obtained by augmenting heart rate data with HF measurement.

%from three sets of predictor variables: 1) heart rate; 2) heat flux; and 3) heart rate and heat flux combined. The results show that the best overall accuracy is obtained by augmenting HR with HF measurement.

\section{Methods}

\subsection{Experimental Setting}
The objective of the experiment was to verify whether a single-site heat flux measurement could provide useful information for energy expenditure estimation. To this end, low-to-medium intensity randomized physical exercise routines were performed in laboratory conditions. During the physical activity, the subjects were equipped with a respiratory calorimeter, which provided the reference data for energy expenditure. For heat flux measurement, each subject wore a custom-made bracelet equipped with heat flux and temperature sensors. The heat flux sensor (greenTEG gSKIN-XM, greenTEG AG, Switzerland) was positioned on the medial side of the user's wrist, above the radial artery. The sensor was attached to a small heat sink, the temperature of which was also recorded during the physical activity. Furthermore, the subjects' heart rate was recorded using an ECG chest strap.
The experiments were conducted on 15 healthy persons (nine male, six female) between 23 and 45 of age (mean$\pm$S.D. $34.7\pm 7.0$ years).

\subsection{Exercise protocol}
Each subject's exercise protocol contained activities in five categories: sitting, standing, treadmill walking, cycling, and arm ergometry. The order of the activities was randomized for each person, with the durations of individual activities ranging from 5 to 45 min. Furthermore, the speed and angle of the treadmill during walking were randomized, as well as the intensities of the arm ergometry exercises. The duration of the exercise protocols ranged from 96 to 163 min, with a total of 33.5 h of physical activity data recorded.

% Esimerkkikuva
\begin{figure}[h]
    \centering
    \includegraphics[width=0.7\columnwidth]{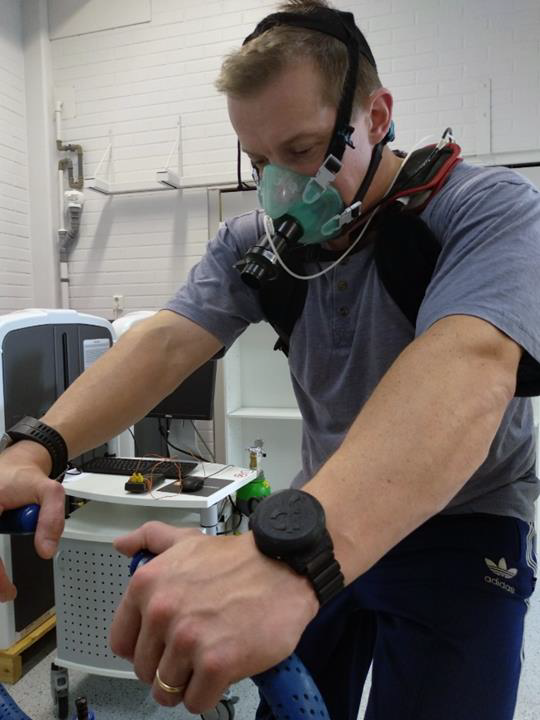}
    \caption{Experimental setup: the subject is wearing a custom-made bracelet equipped with heat flux and temperature sensors on his left hand. A reference EE value was obtained using a respiration calorimeter; furthermore, the subject's heart rate was monitored using an ECG chest strap.}
    \label{fig:heikki}
\end{figure}

\subsection{Data preprocessing and analysis}
% Goal: Estimate the subject's EE based on measurements which can be performed on werable devices.
% --> Goal 2: See wheter a single-site heat flux measurement from the user's wrist makes for a useful addition to other wearable sensors
% Prior work:

The task of estimating the subject's energy expenditure was approached as a regression problem. To this end, the time series data collected from the subjects were composed into a single dataset, with the heart rate (HR) and heat flux (HF) data as predictor variables, and the respiration calorimeter data as the ground truth values $y$ for energy expenditure. The accuracy of the EE estimate was evaluated in three settings: 1) using heart rate data; 2) using heart rate and heat flux data; and 3) using heat flux data.

During the experiments, the heat flux measurements and the heat sink temperatures (T) were sampled at $20\,\text{Hz}$. The ECG equipment and the calorimeter used in this study register each R-R interval and exhaled breath, respectively; these values were composed into 30 s average intervals. Likewise, the heat flux and temperature measurements were downsampled to 30 s intervals. To accommodate the accumulation of heat in the heat sink, the median values of the past 1--2 and 2--8 min of data were used as additional predictor variables for the heat flux and heat sink temperatures. To account for the differences between the physical parameters of the subjects, each person's age, gender, height, weight, and overall physical activity (on the scale from 1 to 10) were considered as exogenous variables for the regression problem. For dimensionality reduction, these five values were projected into a single variable $X_{\text{proj}}$ using Principal Component Analysis (PCA). The predictor variables used in each setting are listed in Table \ref{tab:predictors}.

\begin{table}[h]
    \centering
    \caption{Predictor variables used for EE estimation in three scenarios}
    \begin{tabular}{l c c c}
    \toprule
        Variable            & HR & HR/HF & HF \\ \midrule
        $HR$                  & x    & x       &      \\
        $HF$                  &      & x       & x    \\
        $\text{median}\left( HF_{30-90\text{s}} \right)$  &      & x       & x    \\
        $\text{median}\left(HF_{120-420\text{s}} \right)$ &      & x       & x    \\
        $T$        &      & x       & x    \\
        $\text{median}\left(T_{30-90\text{s}} \right)$  &      & x       & x    \\
        $\text{median}\left(T_{120-420\text{s}} \right)$ &      & x       & x    \\
        $X_{\text{proj}}$        & x    & x       & x    \\
        \bottomrule\\
    \end{tabular}
    \label{tab:predictors}
\end{table}

For the sake of simplicity, the energy expenditure $y$ was modeled as an ordinary least squares problem
\begin{equation}
    y = H \theta + v     \label{eq:model}
\end{equation}
where $H$ is the observation matrix, $\theta$ is a vector of model parameters, and $v$ is a error vector for which $\langle v \rangle= 0$. For the model in \eqref{eq:model}, the parameters $\theta$ were found by minimizing the least squares criterion using the pseudoinverse
\begin{equation}
    \hat{\theta} = \left( H^{\text{T}}H \right)^{-1} H^{\text{T}}y,
\end{equation}
from which the EE estimates were obtained as $\Hat{y}=H\hat{\theta}$.
The least squares model was evaluated in each scenario using the leave-one-out cross validation: the model was fitted into the data from all but one subject, and the performance of the model was tested with the left-out subject, repeating the process for all subjects. The coefficient of determination, $R^2$, was chosen as the performance metric, because it has the mean of EE $\bar{y}$ as the baseline at $R^2=0$
\begin{equation}
    R^2 = 1 - \frac{\text{mean}\left(y-\hat{y}\right)^2}{\text{mean}(y-\bar{y})^2}.
\end{equation}
\section{Results and discussion}
The EE regression model was tested using all three groups of predictor variables (HR, HR/HF and HF) separately. In each case, the validation was performed using both the full dataset and a subset of the data comprising only activities at rest or low intensity (sitting, standing, and walking). The box plots of the cross-validation results in Fig. \ref{fig:boxplots} show that the best performance across both intensity classes is obtained using heart rate data in conjunction with heat flux measurement. While the median $R^2$ values show only a minor increase over the HR-based estimates, the mean values are significantly higher as the lower half of the samples yield higher $R^2$ values than with HR-based estimates.

While the regression based on HF only does not reach the same level of performance as the HR- or HR/HF-based estimates in terms of $R^2$, the mean values are still well above zero. Remarkably, at low intensities, the median $R^2$ values match those of the HR-based estimates, even though there are more outliers and the model is less precise. 

Overall, the results suggest that at low-to-mid activity levels, heat flux is a good predictor of energy expenditure, and will improve the EE estimate when used in conjunction with heart rate information. This is likely due to the thermal radiation and convection being the primary heat transfer mechanisms in the absence of perspiration, in other words, in low-to-mid level physical exercise. On the other hand, heart rate typically fluctuates more at rest and low intensities \cite{HRV}, decreasing its precision as a predictor of energy expenditure.

It can be concluded that a heat flux sensor would be an advantageous feature in a wearable used for daily calorie consumption tracking. However, the HF sensor setup used in the current study does not take into account the effect of evaporative heat transfer, that is, perspiration. A possible solution for this is to use humidity sensors; this is a topic of further study. Moreover, using a nonlinear model for EE estimation would likely result in a higher accuracy in terms of $R^2$, even more so if the model took into account the temporal nature of the measurement data. To this end, the use of models such as recurrent neural networks should be investigated.

%%\begin{table}[htbp]
%%\caption{Cross-validation results in terms of mean $R^2$}
%%\begin{center}
%%\begin{tabular}{l r r}
%%\toprule
%%Inputs & All & Low intensity \\
%%\midrule
%%Heart rate & $0.36$ & $0.30$ \\
%%Heart rate + heat flux & $0.50$ & $0.48$ \\
%%Heat flux & $0.16$ & $-0.22$ \\
%%\bottomrule
%%\end{tabular}
%%\label{tab:taul1}
%%\end{center}
%%\end{table}

\begin{figure}[t]
    \centering
    \includegraphics[width=.95\columnwidth]{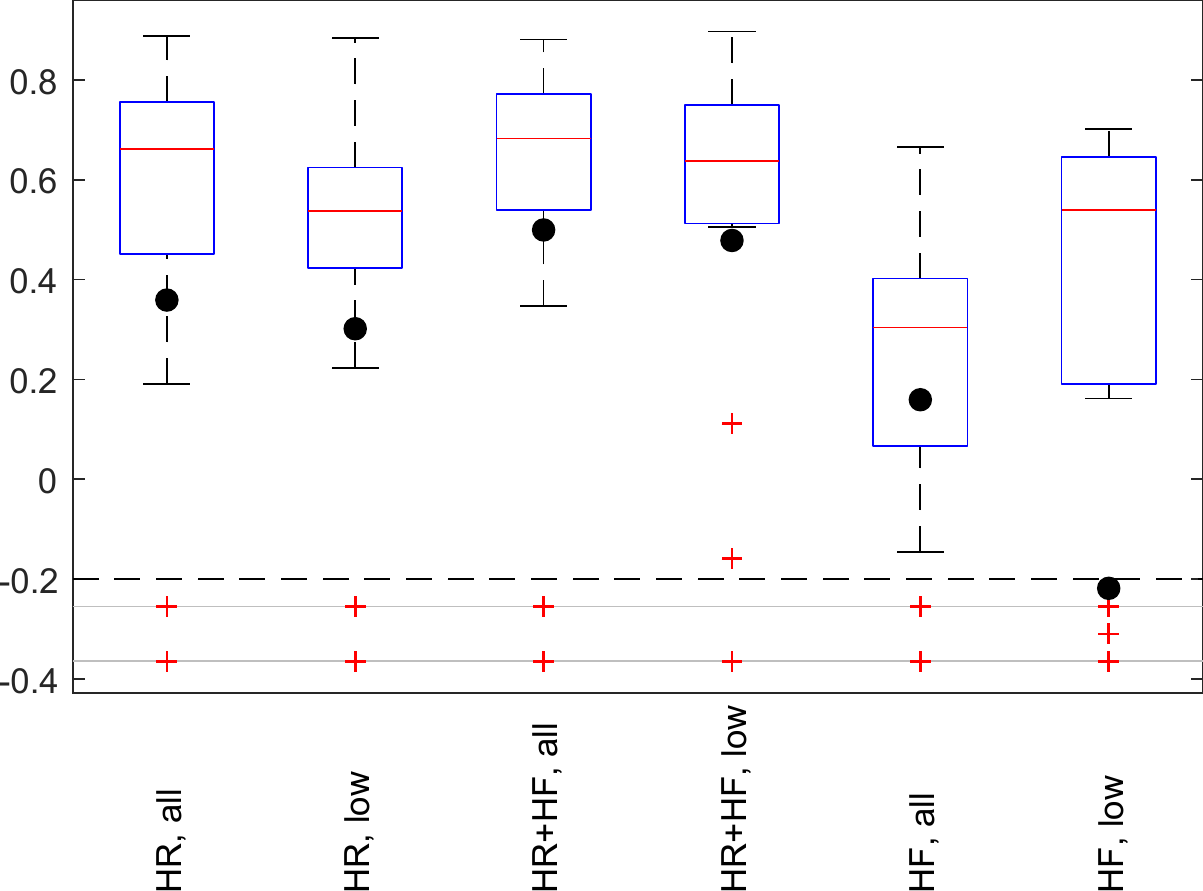}
    \caption{Box plots of the cross-validation results in terms of $R^2$ values. The median value of each group is indicated by a red line. The boxes and whiskers each depict a quartile of the population, with outliers indicated by a red "+" sign. Outliers ranging between $-0.5$ and $-8.5$ are compressed and shown between the gray horizontal lines. The mean of each group is indicated by a black dot.}
    \label{fig:boxplots}
\end{figure}
%\section{Discussion} 
% Yhdistetty results-kappaleen kanssa
\section{Conclusions}
Estimation of low-to-mid intensity energy expenditure was performed on 15 subjects based on heart rate, heat flux, and a combination of these two. The results showed that the best accuracy in terms of $R^2$ values was achieved by augmenting the heart rate measurement with heat flux. Moreover, heat flux appears to be a particularly effective predictor at low intensities or when the subject is at rest. The results suggest that heat flux measurement could be an advantageous feature for wearable devices.
\section*{Acknowledgment}
The authors would like to thank greenTEG AG, Switzerland, for providing the heat flux sensors used in this study. This work was supported by Business Finland (project Q-Health).

% --------------- Lähdeviitteet -----------------------

\bibliographystyle{IEEEtran}
\bibliography{references}

\end{document}